\documentclass[conference]{IEEEtran}

\usepackage{cite}
\usepackage{amsmath,amssymb,amsfonts}
\usepackage{algorithmic}
\usepackage{graphicx}
\usepackage{textcomp}
\usepackage{fancyvrb}
\usepackage{balance}

\begin{document}

\title{Serverless Data Analytics with Flint}

\author{
Youngbin Kim and Jimmy Lin\\
David R. Cheriton School of Computer Science\\
University of Waterloo\\
Waterloo, Ontario, Canada \\
\{youngbin.kim,jimmylin\}@uwaterloo.ca
}

\maketitle

\begin{abstract}
Serverless architectures organized around loosely-coupled function
invocations represent an emerging design for many applications. 
Recent work mostly focuses on user-facing products and
event-driven processing pipelines. In this paper, we explore
a completely different part of the application space and examine the
feasibility of analytical processing on big data using a serverless
architecture. We present Flint, a prototype Spark execution engine
that takes advantage of AWS Lambda to provide a pure pay-as-you-go
cost model. With Flint, a developer uses PySpark exactly as before,
but without needing an actual Spark cluster.
We describe the design, implementation,
and performance of Flint, along with the challenges associated with
serverless analytics.
\end{abstract}

\begin{IEEEkeywords}
serverless computing, cloud computing, data analytics, data science
\end{IEEEkeywords}

\section{Introduction}

Serverless computing~\cite{Baldini_etal_2017,Savage_CACM2018} represents a natural
next step of the ``as a service'' and resource sharing trends in cloud
computing. Specifically, ``function as a service'' offerings such as
AWS Lambda allow developers to write blocks of code with well-defined
entry and exit points, delegating all aspects of execution to the
cloud provider. Typically, these blocks of code are stateless, reading
from and writing to various ``state as a service'' offerings (databases,
message queues, persistent stores, etc.).

Standard serverless deployments are characterized by
asynchronous, loosely-coupled, and event-driven processes that touch relatively small
amounts of data~\cite{Hendrickson_etal_2016}. Consider a canonical
example that Amazon describes:\ an image processing pipeline such that
when the user uploads an image to a website, it is placed in an S3
bucket, which then triggers a Lambda to perform thumbnail
generation. The Lambda may then enqueue a message that triggers
further downstream processing. Most serverless applications are user
facing, even if users are not directly involved in the processing
pipeline.

This paper explores serverless architectures for a completely
different use case:\ large-scale analytical data processing by data
scientists. We describe Flint, a prototype
Spark execution engine that is completely implemented using AWS Lambda
and other services. One key feature is that we
realize a pure pay-as-you-go cost model, in that there are zero costs for
idle capacity. With Flint, the data scientist can transparently use
PySpark without needing an actual Spark cluster, instead paying only
for the cost of running individual programs.

The primary contribution of our work is a demonstration that it is
indeed possible to build an analytical data processing framework using
a serverless architecture. Critically, we accomplish this using cloud
infrastructure that has no idle costs. It is straightforward to see
how workers performing simple ``map'' operations can execute inside
Lambda functions. Physical plans that require data shuffling, however,
are more complicated:\ Flint takes advantage of distributed message
queues to handle shuffling of intermediate data, in effect offloading data movement to
yet another cloud service. 

\section{Background and Design Goals}

Our vision is to provide the data scientist with an experience that is
indistinguishable from ``standard'' Spark.
The only difference is that the user supplies configuration data to
use the Flint serverless backend for execution.
In this context, we explore system performance tradeoffs in
terms of query latency, cost, etc.

Currently, Flint is built on AWS, primarily using Lambda and other
services. All input data to an analytical query are assumed to reside
in an S3 bucket, and we assume that results are written to another S3
bucket or materialized on the client machine. The AWS platform was selected because it remains the
most mature of the alternatives, but in principle Flint can be
re-targeted as other cloud providers have similar offerings.

One major design goal of Flint is to provide a truly pay-as-you-go
cost model with no costs for idle capacity. This needs a bit of
explanation:\ as a concrete example, Amazon Relational Database
Service (RDS) requires the user to pay for database instances (per
hour). This is {\it not} pay as you go because there are ongoing costs even
when the system is idle. Therefore, this
means that one obvious implementation of using RDS to manage intermediate data
would violate our design goal. In general,
we cannot rely on any persistent or long-running daemons.

Note that this is a challenging, but also worthwhile, design goal. In
a cloud-based environment, there are a limited number of options for
Spark analytics. One option is to offload cluster management
completely to a cloud provider via something like AWS EMR, which
starts up a Spark cluster for each user job. The downside is that a
lot of time is wasted in cluster initialization.

\begin{figure*}[t]
\centering\includegraphics[width=0.6\textwidth]{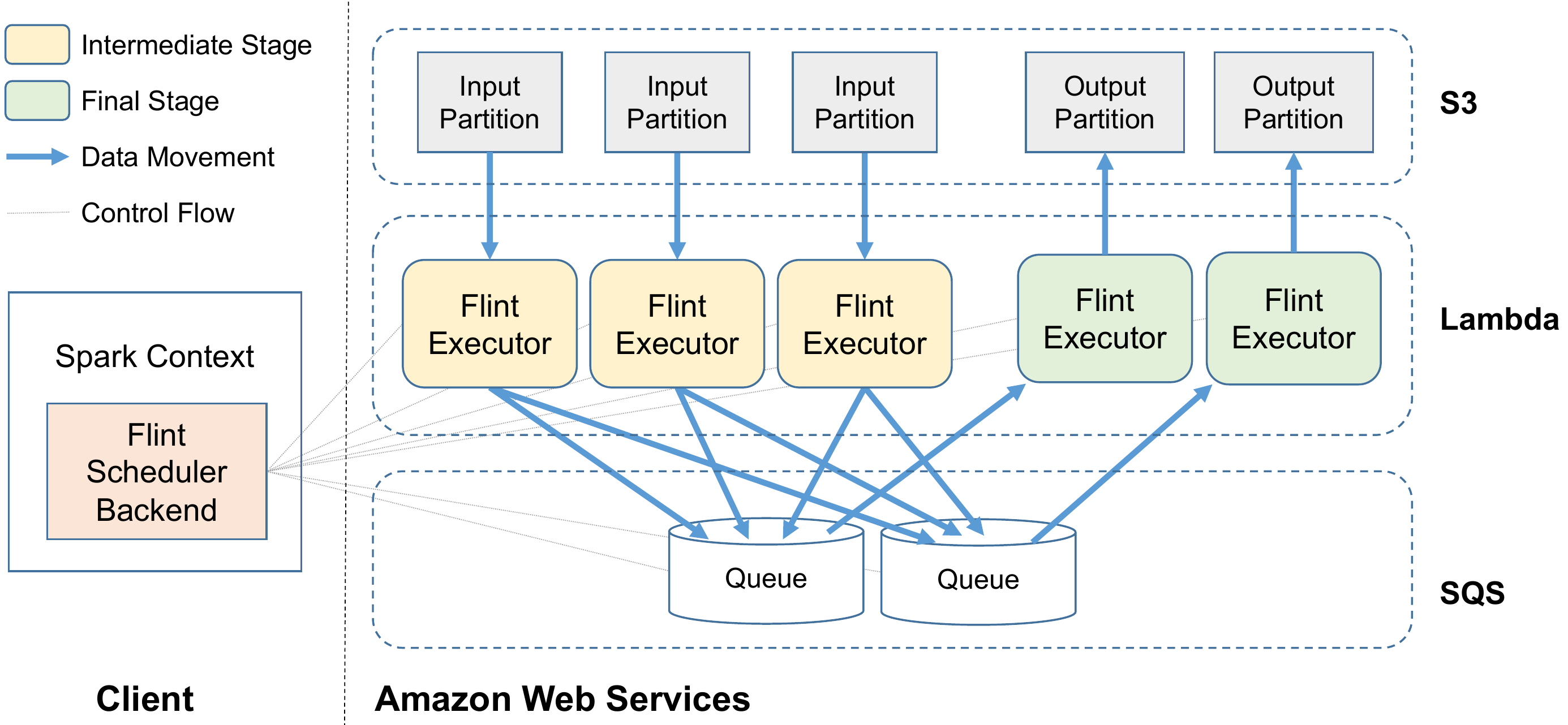}
\caption{Overview of the Flint architecture.}
\label{fig:arch-overview}
\vspace{-0.3cm}
\end{figure*}

The alternative is to manage one's own Spark clusters in the cloud
(on EC2 instances). There are, of course, tools to help with this,
ranging from the UI of Databrick's Unified Analytics Platform to
full-fledged orchestration engines such as Netflix's Genie. Even if
cluster startup and teardown were completely automated (and
instantaneous, let's even say), the fact remains that the organization
pays for cluster instances for the entire time the cluster is up;
charges accumulate even when the cluster is idle. For large
organizations, this is less of an issue as there is more predictable
aggregate load from teams of data scientists, but for smaller
organizations, usage is far more sporadic and difficult to estimate a
priori.

In contrast, we believe that serverless analytics with pay-as-you-go
pricing is compelling, particularly for {\it ad hoc} analytics and
exploratory data analysis.
This is exactly what our Flint prototype provides.


\section{Flint Architecture}

The overall architecture of Flint is shown in
Figure~\ref{fig:arch-overview}. At a high-level, Spark tasks are
executed in AWS Lambda, and intermediate data are held in Amazon's
Simple Queue Service (SQS), which handles the data shuffling necessary
to implement many transformations.

To maximize compatibility with the
broader Spark ecosystem, Flint reuses as many existing Spark
components as possible. When a Spark job is submitted, the sequence of
RDD transformations (i.e., the RDD lineage) is converted into a
physical execution plan using the DAG Scheduler. The physical plan
consists of a number of stages, and within each stage, there is a
collection of tasks. The Task Scheduler is then responsible for
coordinating the execution of these tasks. Spark provides pluggable
execution engines via the \texttt{SchedulerBackend} interface:\ Spark
by default comes with implementations for local mode, Spark on YARN
and Mesos, etc. Flint provides a serverless
implementation of \texttt{SchedulerBackend}; everything else remains
unchanged from standard Spark.
The primary advantage of this design is that we can reuse existing
Spark machinery for query planning and optimization, and Flint only
needs to ``know about'' Spark execution stages and tasks in the
physical plan.

The Flint \texttt{SchedulerBackend} (hereafter ``scheduler''), which lives inside the Spark
context on the client machine, is responsible for
coordinating Flint executors to execute a particular physical plan.
The scheduler receives tasks from Spark's Task
Scheduler, and for each task, our implementation extracts and
serializes the information that is needed by the Flint executors.
This information includes the serialized code to execute,
metadata about the relationship of this task to the entire physical
plan, and metadata about where the executor reads its input and writes
its output. When this serialization is complete, the scheduler
asynchronously launches the Flint executors
on AWS Lambda, with the serialized task as part of the request. After
a Flint executor has completed its task, the
scheduler processes the response. Once all
tasks of the current stage complete, executors for tasks of
the next stage are launched, repeating until the entire physical plan has
been executed.

\subsection{Flint Executor} 

A Flint executor is a process running inside an Amazon Lambda function
that executes a task in a Spark physical plan. Since the startup latency of a Lambda
invocation is small once the function has been ``warmed up'' (more
discussion later), each Lambda instance only processes a single
task. This is different from standard Spark,
where executors are long-running processes.

Once a Flint executor is initialized, it first deserializes the task
information from the request arguments. 
From the input partition metadata, the executor creates an input
iterator to read from the appropriate input partition. For the first
stage in a plan, this iterator will fetch a range of bytes from an S3
object. For most other stages, the input iterator will fetch from a
designated SQS queue (discussed in detail below).

Once the input iterator is ready, it is passed to the deserialized
function (i.e., code to execute) from the task as an argument; this yields the output iterator.
If the task is in the final (result)
stage of the execution plan, there are two possibilities:\ If the
final action on the RDD is \texttt{saveAsTextFile}, outputs are
materialized to another S3 bucket; otherwise, the results are
materialized within the executor and passed back to the
scheduler (for example, if the data scientist
calls the \texttt{collect} action).

When a task is part of an intermediate stage, the execution plan
requires the output to be shuffled so that all values for a key
are placed in the same partition. The shuffling is part
of the physical plan created by Spark; the Flint executors simply
execute the task, and thus are not explicitly aware of the actual
RDD transformation (e.g., if the shuffling is part of
\texttt{reduceByKey} or \texttt{join}, etc.). Since the execution time
of a Lambda invocation has a limit of 300 seconds, it is not possible
to guarantee that the Flint executors from the previous stage are
still alive to pass data to executors running tasks
from the next stage. Thus, we need some external data store to deliver
the intermediate output. Flint uses Amazon's Simple Queue Service
(SQS) for this purpose.

Once an executor of a task belonging to an intermediate stage has
computed the output iterator, the hash partition function 
(or custom partition function if specified) is used to
decide which partition each output object will be assigned to. 
The executor groups objects by the destination partition in memory.
However, if memory usage becomes too high during this process, the
executor flushes its in-memory buffers by creating a batch of SQS messages and
sending them to the appropriate queue for each partition. After all
output data are sent to SQS queues, the executor terminates and returns
a response containing a variety of diagnostic information (e.g.,
number of messages, SQS calls, etc.).

Once all tasks of the current stage are completed, executors for tasks
in the next stage will be launched. These executors read from their
corresponding SQS queues and aggregate data in memory.
Results are passed to the iterator of the function associated with the task,
as described earlier. Since we are using in-memory data
structure for aggregation, memory forms a bottleneck. Due to
the complexities of implementing on-disk multi-pass aggregation
algorithms in the Lambda environment, we currently address this
problem by increasing the number of partitions such that we do not
overflow memory. This solution appears to be adequate, since it
takes advantage of the elasticity provided by AWS Lambda.

Queue management is performed by the scheduler. Before the execution
of each stage, the scheduler initializes the necessary partitions.
Partition metadata (i.e., source and destination queues) are passed as
part of the Lambda request. The scheduler also handles cleanup.

\subsection{Overcoming Lambda Limitations}

The current Flint implementation supports PySpark, which
counter-intuitively is easier to support than Scala Spark. The Flint
\texttt{SchedulerBackend} on the client is implemented in Java,
but the Flint executors in AWS Lambda are implemented in Python.
This design addresses one of the biggest issues with AWS
Lambda:\ the long cold startup time of function invocations. The
first time that a Lambda is invoked (or after some idle period), AWS
needs to provision the appropriate runtime container. Java Lambdas
tend to have large deployment packages due to numerous dependencies,
whereas Python Lambdas are significantly more lightweight; thus, they
start up faster. Furthermore, in the default Spark executor
implementation, to run PySpark code, data (from S3) is first read in
the JVM and then passed to a Python subprocess using pipes. In Flint,
we bypass this extra wrapper layer, and Python code is able to read
from S3 directly. As we later show, this has significant performance
advantages.

Another limitation of AWS Lambda is that execution duration per
request is capped at 300 seconds. This leads to the failure of
long-running tasks. In order to avoid this problem, we chain
executors:\ if the running time has almost reached the limit, the
Flint executor stops ingesting new input records. Then, the current
state, including how much of the input split has been read, is
serialized and returned to the scheduler, which
launches a new Flint executor to continue processing the uncompleted
input split from where the previous invocation left off. Since the
function is already warm, the cost of using chained executors is
relatively low.

A third limitation of Lambda comes from a number of resource
constraints. Each invocation is limited to a maximum memory allocation
of 3008 MB. Thus, it is important for the Flint executor to carefully
manage in-memory data. There is a limitation of 6 MB on the size of the
request payload for an invocation. This payload is used to hold the
serialized task data, which is typically much smaller. However, for larger tasks we are
currently implementing a workaround for this size restriction by
splitting the payload into smaller pieces. These can be uploaded to
S3, and the scheduler can direct the Lambda functions to fetch the
relevant data to complete initialization.

\section{Experimental Evaluation}

\begin{table*}[t]
\begin{center}
\begin{tabular}{|l|p{2cm}|p{1.25cm}|p{1.25cm}||p{1.25cm}|p{1.25cm}|p{1.25cm}|}
\hline
 & \multicolumn{3}{c||}{Query Latency (s)} & \multicolumn{3}{|c|}{Estimated Cost (USD)} \\
\cline{2-7}
\textbf{} & \textbf{Flint} & \textbf{PySpark}  & \textbf{Spark}  & \textbf{Flint} & \textbf{PySpark}  & \textbf{Spark} \\
\hline
\hline
0 & 101 [93 - 109] & 211 & 188 & 0.20 & 0.41 & 0.37  \\
1 & 190 [186 - 197] & 316 & 189 & 0.59 & 0.61 & 0.37 \\
2 & 203 [201 - 205] & 314 & 187 & 0.68 & 0.61 & 0.36 \\
3 & 165 [161 - 169] & 312 & 188 & 0.48  & 0.61 & 0.36 \\
4 & 132 [122 - 142] & 225 & 189 & 0.33 & 0.44 & 0.37 \\
5 & 159 [142 - 177] & 312 & 189 & 0.45 & 0.60 & 0.37 \\
6 & 277 [272 - 281] & 337 & 191 & 0.56 & 0.66 & 0.37  \\
\hline
\end{tabular}
\vspace{0.2cm}
\caption{Query latency and cost comparisons.}
\label{tab:results}
\vspace{-0.7cm}
\end{center}
\end{table*}


We evaluated Flint by comparing its performance with a Spark cluster
running the Databricks Unified Analytics Platform. The entire cluster
comprises 11 m4.2xlarge instances (one driver and ten workers), with a total of 80 vCores (Amazon's
computation unit) of processing capacity. For AWS Lambda, we allocated
the maximum amount of memory possible, which is 3008 MB. The developer
can also configure the maximum number of concurrent invocations; we
set this to 80 to match the Spark cluster, under the assumption that
one Lambda invocation roughly uses one vCore. AWS is not completely
transparent about the instances running AWS Lambda; documentation
refers to a ``general purpose Amazon EC2 instance type, such as an M3 type'' without
additional details. Thus, this is
the best that we can do to ensure that all conditions consume the same
hardware resources. In all cases, we used the latest version of the Databricks runtime,
which is based on Spark 2.3.

Our evaluations examined three different conditions:\ PySpark code
running on Flint, PySpark code running on the Spark cluster, and
equivalent Scala Spark code running on the Spark cluster. For the
Spark cluster, we only measure query execution time (derived from
the cluster logs) and do not include startup costs of the cluster (around
five minutes). This puts Spark performance in the best possible
light. We had separately examined Amazon EMR, which initializes
clusters automatically per job---but for reasons unknown from
available documentation, its performance (even excluding startup
costs) was significantly worse than a Spark cluster we provisioned
ourselves.

For evaluation, we considered a typical exploratory data analysis task
described in a popular blog post by Todd Schneider~\cite{Blog}. The
New York City Taxi \& Limousine Commission has released a detailed
historical dataset covering approximately 1.3 billion taxi trips in
the city from January 2009 through June 2016. The entire dataset is
stored on S3 and is around 215 GB. Each record contains information
about pick-up and drop-off date/time, trip distance, payment type,
tip amount, etc.
Inspired by Schneider's blog post, we evaluated the following queries,
some of which replicate his analyses:

\smallskip \noindent {\bf Q0.} Line count. In this query, we
simply counts the number of lines in the dataset. This evaluates the
raw I/O performance of S3 under our experimental conditions.

\smallskip \noindent {\bf Q1.} Taxi drop-offs at Goldman Sachs
headquarters, 200 West St. This query filters by geo coordinates and
aggregates by hour, as follows:

\smallskip
\begin{Verbatim}[fontsize=\small]
  arr = src.map(lambda x: x.split(',')) \
    .filter(lambda x: inside(x, goldman)) \
    .map(lambda x: (get_hour(x[2]), 1)) \
    .reduceByKey(add, 30) \
    .collect()
\end{Verbatim}
\smallskip

\noindent This is exactly the query issued in PySpark to both Flint
and the Spark cluster. The Scala Spark condition evaluates exactly the
same query, except in Scala. Note that Flint is able to support UDFs
transparently.

For brevity, we omit code for the following queries and instead
provide a concise verbal description.

\smallskip \noindent {\bf Q2.} Same query as above, but for
Citigroup headquarters, at 388 Greenwich St. 

\smallskip \noindent {\bf Q3.} Goldman Sachs taxi drop-offs with tips
greater than \$10. Who are the generous tippers?

\smallskip \noindent {\bf Q4.} Cash vs.\ credit card payments. This
query computes the proportion of rides paid for using credit cards,
aggregated monthly across the dataset.

\smallskip \noindent {\bf Q5.} Yellow taxi vs.\ green taxi. This query
computes the number of different taxi rides, aggregated by month.

\smallskip \noindent {\bf Q6.} Effect of precipitation on taxi trips,
i.e., do people take the taxi more when it rains? This query aggregates
rides for different amounts of precipitation.

\medskip \noindent Table~\ref{tab:results} reports latency (in seconds) and
estimated cost of each query (in USD) under the three different
experimental conditions discussed above. For Flint, we
report averages over five trials (after warm-up) and show 95\% confidence intervals in
brackets. The latency of PySpark and Spark exhibit little variance,
and thus we omit confidence intervals (over three trials) for brevity.
Estimated costs for Spark and PySpark are computed
as the query latency multiplied by the per-second cost of
the cluster. For Flint, we used logging information to
compute the execution duration of the AWS Lambdas and the associated
SQS costs.

We find that latency is roughly the same for all queries on Spark and
appears to be dominated by the cost of reading from S3. This is
perhaps not surprising since none of our test queries are shuffle
intensive, as the number of intermediate groups is
modest. Interestingly, for some queries, Flint is actually faster than
Spark. The explanation for this can be found in Q0, which simply
counts lines in the dataset and represents a test of read throughput
from S3. Evidently, the Python library that we use (boto) achieves
much better throughput than the library that Spark uses to read from
S3. This is confirmed via microbenchmarks that isolate read throughput
from a single EC2 instance. In our queries, the
performance of Flint appears to be dependent on the
number of intermediate groups, and this variability makes sense as
we are offloading data movement to SQS. PySpark
is much slower than Flint because every input record passes from the
JVM to the Python interpreter, which adds a significant amount of
overhead. In terms of query costs, Flint is in general more expensive
than Spark, even for queries with similar running times (Flint has
additional SQS costs). Although a direct cost conversion between
Lambda and dedicated EC2 instances is difficult (the actual instance
type and the multiplexing mechanism are both unknown), it makes sense that
Lambda has a higher per-unit resource cost, which corresponds to the
price we pay for on-demand invocation, elasticity, etc.

For the above reasons, it is difficult to obtain a truly fair
comparison between Flint and Spark. Nevertheless, our experiments show
that serverless analytics is feasible, though a broader range of
queries is needed to tease apart performance and cost
differences---for example, large complex joins, iterative algorithms,
etc.
However, results do suggest that data shuffling is a
potential area for future improvement.

\section{Related Work}

The origins of Flint can be traced back to a course project at the
University of Waterloo in the Fall of 2016. Since then, there
have been other attempts at exploring serverless architectures for
data analytics. In June 2017, Databricks announced a serverless
product~\cite{DatabricksServerless}, best described as a more
flexible resource manager:\ administrators define a ``serverless
pool'' that elastically scales up and down. This can be viewed as more
convenient tooling around traditional Spark clusters, and is not
serverless in the sense that we mean here.

In November 2017, Qubole announced an implementation of Spark on AWS
Lambda~\cite{QuboleServerless}. This effort shares the same goals as Flint,
but with several important differences. Qubole
attempted to ``port'' the existing Spark executor infrastructure onto
AWS Lambda, whereas Flint is a from-scratch implementation. As a
result, we are better able to optimize for the unique execution
environment of Lambda. For example, Qubole reports executor startup time to be around two minutes in the cold start case.
In addition, Qubole's implementation uses S3 directly for
shuffling intermediate data, which differs from our SQS-based shuffle.
Using S3 allows Qubole's executors to remain more faithful
to Spark, but we believe that the I/O patterns
are not a good fit for S3.



Amazon provides two data analytics services that are worth
discussing:\ Amazon Athena (announced November 2016) and Amazon
Redshift Spectrum (announced July 2017). Both are targeted at more
traditional data warehousing applications and only support SQL queries,
as opposed to a general-purpose computing platform
like Spark. Athena offers a pay-as-you-go, per-query pricing with zero
idle costs, similar to Flint, but under the covers it uses the Presto
distributed SQL engine for query execution, so architecturally it is not serverless.
Redshift Spectrum is best described as
a connector that supports querying over S3 data; the customer still
pays for the cost of running the instances that comprise the Redshift
cluster itself (i.e., per hour charge, even when idle).

PyWren~\cite{Jonas_etal_SoCC2017} is another project advocating a
serverless execution model for analytics tasks, although unlike our
effort PyWren does not attempt to target Spark or any specific analytics
framework. Since Flint is a Spark execution engine, it supports
arbitrary RDD transformations; in contrast, PyWren examines only
three classes of dataflow patterns:\ map-only, map + monolithic
reduce, and MapReduce using either S3 or Redis for shuffling (the
latter is not pay as you go).

Serverless computing in general is an emerging computing paradigm and
previous work has mostly focused on examining system-level issues in
{\it building} serverless infrastructure~\cite{McGrath_Brenner_2017}
as opposed to designing applications. Indeed, as Baldini et
al.~\cite{Baldini_etal_2017} write, the advantages of serverless
architectures are most apparent for bursty, compute-intensive
workloads and event-based processing pipelines. Data analytics
represents a completely different workload and Flint opens up
exploration in a completely different part of the application space.

\section{Future Work and Conclusions}

There are a number of future directions that we are actively
exploring. We have not been able to conduct an experimental evaluation
between Qubole's implementation and Flint, but the design choice of
using S3 vs.\ SQS for data shuffling should be examined in
detail. Each service has its strengths and weaknesses, and we can
imagine hybrid techniques that exploit the strengths of both.

Robustness is an issue that we have not explored at all in this work, although to
some extent the point of serverless designs is to offload these
problems to the cloud provider. Executor failures can be overcome by
retries, but another issue is the at-least-once message semantics of
SQS. Under typical operating circumstances, SQS messages are only
delivered once, but AWS documentation explicitly acknowledges the
possibility of duplicate messages. We believe that this issue can be
overcome with sequence ids to deduplicate message batches, as the
exact physical plan is known ahead of time.

Another ongoing effort is to ensure that higher-level Spark libraries
(e.g., MLlib, SparkSQL, etc.)~work with Flint. To the extent that
\texttt{SchedulerBackend} provides a clean abstraction, in theory
everything should work transparently. However, as every developer
knows, abstractions are always leaky, with hidden dependencies. We
are pushing the limits of our current implementation by iteratively 
expanding the scope of Spark libraries and features that we use.

To conclude, we note that Flint is interesting in two different
ways:\ First, we show that big data analytics is feasible using a
serverless architecture, and that we can coordinate the data shuffling
associated with analytical queries in a restrictive execution
environment. Second, there are compelling reasons to prefer using our
execution engine over Spark's default, particularly for {\it ad hoc}
analytics and exploratory data analysis:\ the tradeoff is a bit of
performance for elasticity in a pure pay-as-you-go cost model. Thus,
Flint appears to be both architecturally interesting as well as
potentially useful.

\section{Acknowledgments}

Flint traces back to a serverless analytics
prototype called Iris developed as part of a course project at the
University of Waterloo in Fall 2016 with Jonathan Ma, Ronak Patel, and
Pranjal Shah. Although Flint does not share any code with Iris, we'd
like to acknowledge these individuals for their contributions to early
developments of the serverless concept.



\end{document}